\documentclass[aps,prb,twocolumn,showpacs,preprintnumbers,amsmath,amssymb,superscriptaddress]{revtex4-1}

\usepackage{graphicx}% Include  files
\usepackage{amssymb}
\usepackage{epstopdf}
\usepackage{dcolumn}% Align table columns on decimal point
\usepackage{bm}% bold math
\usepackage{amsmath}
\usepackage{mathrsfs}
\usepackage{siunitx}
\usepackage{subfigure}
\usepackage{physics}
\usepackage{flafter}
\usepackage{caption}
\usepackage{tabularx}
\usepackage{float}
\usepackage{braket,mleftright}
\usepackage[font={small}]{caption}
\newcolumntype{C}[1]{>{\centering\let\newline\\\arraybackslash\hspace{0pt}}m{#1}}
\newcolumntype{P}[1]{>{\centering\arraybackslash}p{#1}}

\newcommand{\ep}{\epsilon}

\newcommand{\vep}{\varepsilon}
\newcommand{\sE}{{\mathcal{E}}}
\newcommand{\fJVm}{{fJV$^{-1}$m$^{-1}$}}

\newcommand{\nCr}{{\bar{n}_{\rm Cr}}}
\newcommand{\anCr}{{\alpha_{\bar{n}_{\rm Cr}}}}
\begin{document}

\title{Effects of filling, strain, and electric field on the 
N\'{e}el vector in antiferromagnetic CrSb} 

\author{In Jun Park}
\email{ipark008@ucr.edu}
\affiliation{Laboratory for Terascale and Terahertz Electronics (LATTE), Department of Electrical and Computer Engineering, University of California, Riverside, CA 92521, USA}

\author{Sohee Kwon}
%\email{skwon054@ucr.edu}
\affiliation{Laboratory for Terascale and Terahertz Electronics (LATTE), Department of Electrical and Computer Engineering, University of California, Riverside, CA 92521, USA}

\author{Roger K. Lake}
\email{rlake@ece.ucr.edu}
\affiliation{Laboratory for Terascale and Terahertz Electronics (LATTE), Department of Electrical and Computer Engineering, University of California, Riverside, CA 92521, USA}
%\affiliation{Center of Spins and Heat in Nanoscale Electronic systems, University of California, Riverside, CA 92521, USA}

\date{\today}

\begin{abstract}
CrSb is a layered antiferromagnet (AFM) with perpendicular magnetic anisotropy, a high N\'{e}el temperature, 
and large spin-orbit coupling (SOC),
which makes it interesting for AFM spintronic applications.
To elucidate the various mechanisms of N\'{e}el vector control,
the effects of strain, band filling, and electric field on the magnetic anisotropy energy (MAE) of bulk and thin-film CrSb are determined
and analysed using density functional theory.
The MAE of the bulk crystal is large (1.2 meV per unit cell).
Due to the significant ionic nature of the Cr-Sb bond, finite slabs are strongly affected
by end termination. 
Truncation of the bulk crystal to a thin film with one surface terminated with Cr and the other 
surface terminated with Sb breaks inversion symmetry,
creates a large charge dipole and average electric field across the film,
and breaks spin degeneracy, such that the thin film
becomes a ferrimagnet.
The MAE is reduced such that its sign can be switched with realistic strain, and
the large SOC gives rise to an intrinsic voltage controlled magnetic anisotropy (VCMA). 
A slab terminated on both faces with Cr remains a compensated AFM, but with the compensation
occurring nonlocally between mirror symmetric Cr pairs.
In-plane alignment of the moments is preferred, 
the magnitude of the MAE remains large, similar to that of the bulk, and it is relatively insensitive to filling.
\end{abstract}

\maketitle %\maketitle must follow title, authors, abstract and \pacs

%%%%%%%%%%%neeed to change the introduction%%%%%%%%%%%%%%%
\section{Introduction}
\begin{figure}[h]
\centering
\includegraphics[width=.8\linewidth]{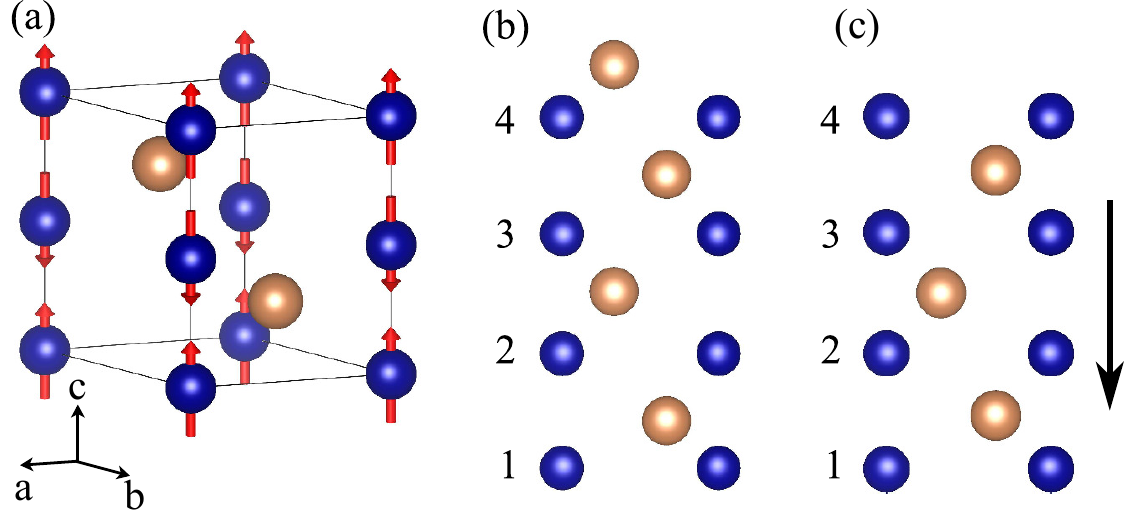}
\caption{\label{fig:structure} 
(a) Bulk antiferromagnetic CrSb crystal structure and spin texture in the ground state.
Blue and brown spheres indicate the Cr and Sb atoms, respectively.
(b) 1.1 nm thin film with a thickness of 2 unit cells. 
(c) The same thin film as in (b) but with the top Sb layer removed.
The numbers
index the Cr atoms, and the arrow indicates the direction of 
positive applied external electric field for VCMA calculations.
}
\end{figure}

Antiferromagnetic (AFM) materials are of great interest for future spintronics applications\cite{2018_Tserkovnyak_RMP}.
Their resonant frequencies are much higher than those of ferromagnetic (FM) materials, 
which allows them to be used in the THz applications \cite{AFM_spintronics_Jungwirth_NNano16, gomonay2014spintronics, keffer1952theory}
and ultrafast switching \cite{lopez2019picosecond}.
However, it is challenging to control and detect the antiferromagnetic states.
There are several methods to control the spins in AFMs such as via exchange bias with a proximity FM layer\cite{nogues1999exchange} 
and the use of electric current by N\'eel spin-orbit torque \cite{wadley2016electrical}.
The latter method has been extensively studied, 
although the results have recently been questioned \cite{No_sot_1,No_sot_2}.
Controlling the N\'eel vector without electric current is promising for ultra low power applications, 
since it has been predicted that magnetization reversal can be achieved with atto joule (aJ) level energy consumption \cite{semenov2017currentless}.
Electric field control of the magnetic properties of AFMs can be realized indirectly through the mechanism of mechanical strain 
created from a piezoelectric substrate \cite{barra2018voltage, yan2019piezoelectric, Mn_lambda, chen2019electric,Mn3Sn_ActaMaterialia19,Mn3NiN_ApplMatInt18,Mn3Pt_NatElect18}
or a combination of strain plus exchange spring \cite{strain_cntl_FeMn_AdvMat20}.
It can also be realized directly through the mechanism of voltage controlled magnetic anisotropy (VCMA). 
This mechanism has been experimentally and theoretically studied for FMs
\cite{2008_VCMA_Tsymbal_PRL,
2009_Giant_VCMA_Nakamura_PRL,
VCMA_die,
VCMA_KWang_SPIN12,
Alzate_Temp_VCMA_APL14,
nozaki2014magnetization,
VCMA_Kiousis_Wang_Carman_PRB15,
skowronski2015underlayer,
skowronski2015perpendicular,
hibino2015electric,
VCMA_Suzuki_PRAppl16,
shiota2017reduction,
Tsymbal_VCMA_PRB17,
VCMA_Exp_SciRep17,
2017_VCMA_Cr_Fe_MgO_ScRep,
kawabe2017electric,
li2017enhancement,
kato2018giant,
SKwon_VCMA_cap_PRAppl19,
SKwon_colossal_VCMA_PRB19,
nozaki2020voltage,
2020_VCMA_Origin_Cr_Fe_MgO_Chen_PRB},
and the experimental results have been recently reviewed.\cite{VCMA_review_Miwa_JPD18}
Technological applications have been described and analysed
\cite{VCMA_RAM_KWang_TNANO15,
VCMA_MRAM_TNANO17,
VCMA_Flip_Flop_IEEEMagLett16}. 
More recently, several theoretical studies of VCMA in the AFM materials FeRh, MnPd, and MnPt have been reported
\cite{zheng2017electric, lopez2019picosecond, su2020voltage, chang2020voltage}.
CrSb crystallizes in the hexagonal NiAs-type structure, and the spins on the 
Cr atoms couple ferromagnetically within the hexagonal plane and antiferromagnetically 
along the hexagonal axis as shown in Fig. \ref{fig:structure}(a).
In the ground state, the N\'eel vector aligns along the hexagonal axis ([0001] direction),
so that it has perpendicular magnetic anisotropy (PMA).
The bands near the Fermi energy are composed of the d-orbitals of the Cr atoms, and
these bands give rise to a large peak in the density of states near the Fermi 
energy.\cite{E-k_CrX_JPCM89,Elec_Struct_Mag_Props_CrX_JMMM07,Mag_Props_JAP07,Struct_Mag_Props_JPCM12}
The Sb atoms provide significant SOC.
CrSb has a high N\'eel temperature (705 K) making it suitable for on-chip applications \cite{takei1963magnetic}.
Recently, CrSb has been used to control the magnetic textures and tune the surface states of topological 
insulators \cite{he2017tailoring,CrSb_on_TI_KWang_PRL18,CrSb_SbTe_JJAP19}.
We examine three different physical mechanisms that alter the magnetic anisotropy 
of bulk and thin-film CrSb:
(i) strain, (ii) electron filling, and (iii) electric field. 
Density functional theory (DFT) calculations of
the magnetostriction coefficient, strain coefficient, filling coefficient, and VCMA coefficient 
characterize the effectiveness of the three methods in modifying the MAE.

\section{Methods}
\label{sec:Method}
We perform first principles calculations as implemented in the Vienna Ab initio Simulation Package 
(VASP) \cite{kresse1993ab} to investigate the effects of strain, electric field, 
and band filling on the magnetic anisotropy of CrSb. 
Projector augmented-wave (PAW) potentials \cite{blochl1994projector} and the generalized gradient approximation (GGA) 
parameterized by Perdew-Burke-Ernzerhof (PBE) are employed \cite{perdew1996generalized}.
A cut-off energy of 500 eV and 8 $\times$ 8 $\times$ 8 $\Gamma$-centered k-point grid were used to make sure the total energy 
converged within $10^{-7}$ eV per unit cell.
A GGA+$U$ implementation was also used to reproduce the magnetic moment on the Cr atom corresponding to the experimental values \cite{CrSb_magmom}.
We used $U_{eff} = 0.25$ eV for the Cr atom where $U_{eff} = U - J$.
From the initial bulk structure, uniaxial strain along the $x$-axis is applied 
and the structure is fully relaxed along $y$ and $z$ axes until all forces on each atom are less than $10^{-3}$ eV\AA$^{-1}$.
Here, the
$x$ and $z$ axes are parallel to the $a$ and $c$ lattice vectors of the hexagonal unit cell shown 
in the Fig. \ref{fig:structure}.
The strain is defined as $\vep = (a - a_0)/a_0 \times 100 \%$ where 
$a$ and $a_0$ are the lattice constants along $x$ with and without strain, respectivley.
The calculated lattice constants without strain are $a_0=4.189$ \AA \space and $c_0=5.394$ \AA, 
which are close to those from experiment \cite{CrSb_magmom}.

To obtain the charge density, a spin-polarized self-consistent calculation is 
performed with the relaxed structure for each strain.
Using the obtained charge densities, 
$E_{\parallel}$ and $E_{\perp}$,
are calculated in the presence of SOC where 
$E_{\parallel}$ and $E_{\perp}$ 
are the total energies per unit cell 
with the N\'eel vector along [1000] and [0001] directions, respectively.
The magnetic anisotropy energy (MAE) is defined as
$E_{\rm MAE} = E_{\parallel}-E_{\perp}$.
For uniformity of comparison between bulk and thin-film structures,
all values of $E_{\rm MAE}$ are reported per bulk unit cell (u.c.) (i.e. per two Cr atoms). 
For MAE calculations, a denser k-point grid (16 $\times$ 16 $\times$ 16) is used for accuracy.
The same procedures are performed to investigate the effect of electron filling on the MAE, 
and the structures are optimized for each number of electrons in the unit cell.

Charge transfer between the Cr and Sb ions is analyzed by calculating both the Bader charges \cite{henkelman2006fast}
and the planar averaged volumetric charge densities \cite{wang2020vaspkit}.
The ``net electronic charge'' on each atom is defined as the number of valence electrons for a given
atom minus the Bader charge on the atom in units of $|e|$.
Thus, a depletion of electrons is a positive electronic charge.
The Bader charges are used to understand the effect of truncation of the bulk to a slab and the application
of an electric field.
For thin-films,
the planar averaged volumetric charge densities at different electric fields are 
calculated by averaging the three-dimensional charge density over the $x-y$ plane for
fixed positions $z$ on the $c$ axis. 

To investigate the effect of strain and electric field on the MAE of 
thin-film CrSb, 
we consider slab structures consisting of 2 and 3 unit cells along the $c$-axis 
($\sim$ 1.1 nm and 1.6 nm).
A 15 {\AA} vacuum layer is included.
The stability of two different surface terminations is quantified by calculating the cohesive
energy defined as $E_{\rm coh} = (E_{\rm crystal} - E_{\rm isolated})/N$ where 
$N$ is the total number of atoms, $E_{\rm crystal}$ is the total energy of the relaxed slab structure,
and $E_{\rm isolated}$ is the sum of the energies of the individual atoms.
For the thin-film structures, a 23 $\times$ 23 $\times$ 1 $\Gamma$-centered 
k-point grid with a 500 eV cutoff energy
is used to ensure the same convergence criteria as the bulk structure.
The structures are fully relaxed until all forces on each atom are less than 
$10^{-3}$ eV\AA$^{-1}$ without changing the volume.
Vertical external electric fields are applied to the slab by introducing a dipole layer in
the middle of the vacuum layer.
The dipole layer also corrects for the built in dipole moment in the CrSb slab structures
to prevent interactions between the artificial periodic images \cite{dipole1_PRB,dipole2_PRB}.
The equilibrium charge density is obtained by performing a spin-polarized 
self-consistent calculation without the electric field.
Then, the charge densities with increasing applied electric fields along c-axis are obtained 
by relaxing the charges from the calculation with the previous electric field.
For each electric field, the MAE is calculated using a $46\times46\times 1$ k-point grid.
\begin{figure}[H]
\includegraphics[width=.95\linewidth]{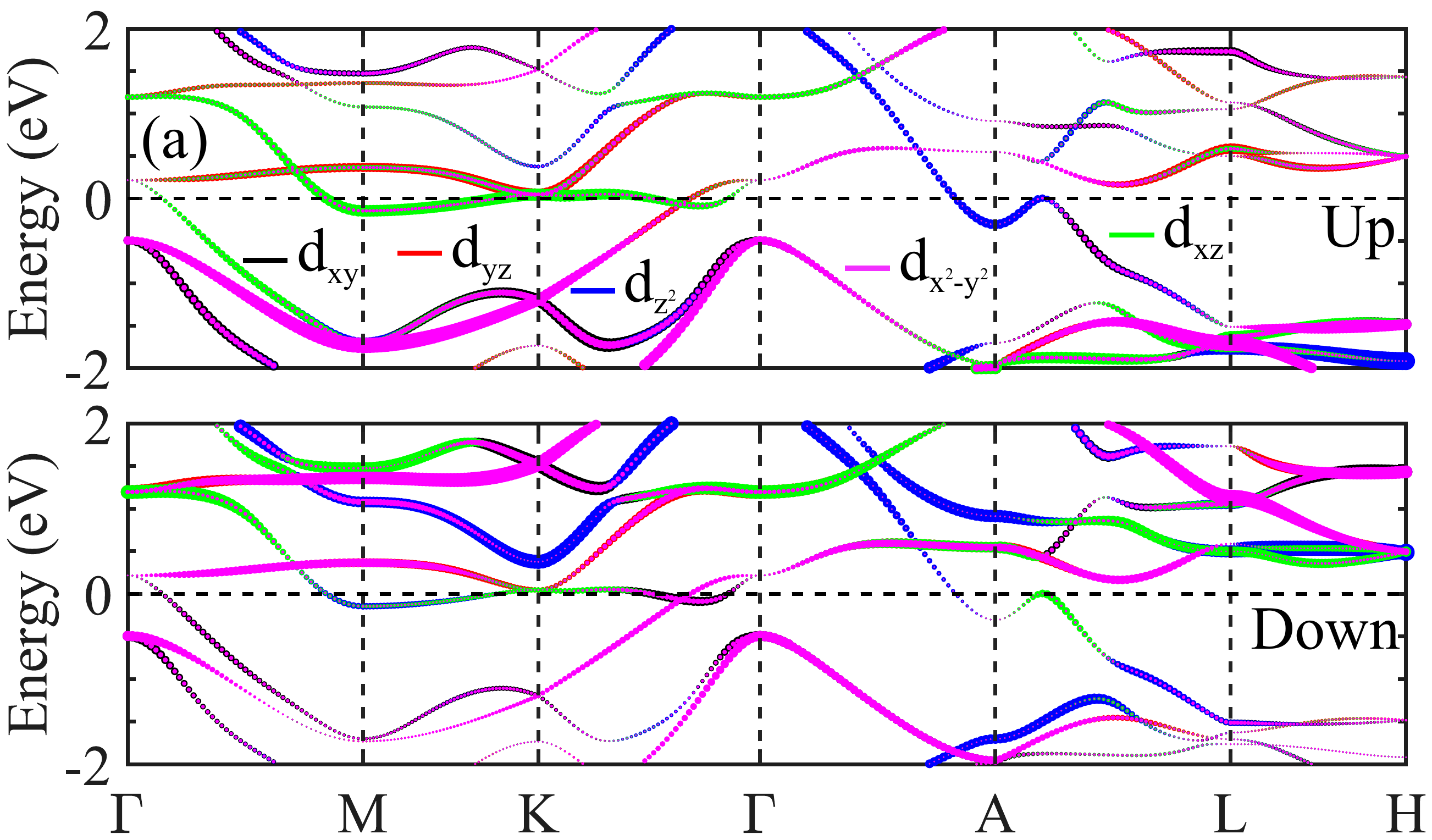}
\includegraphics[width=.95\linewidth]{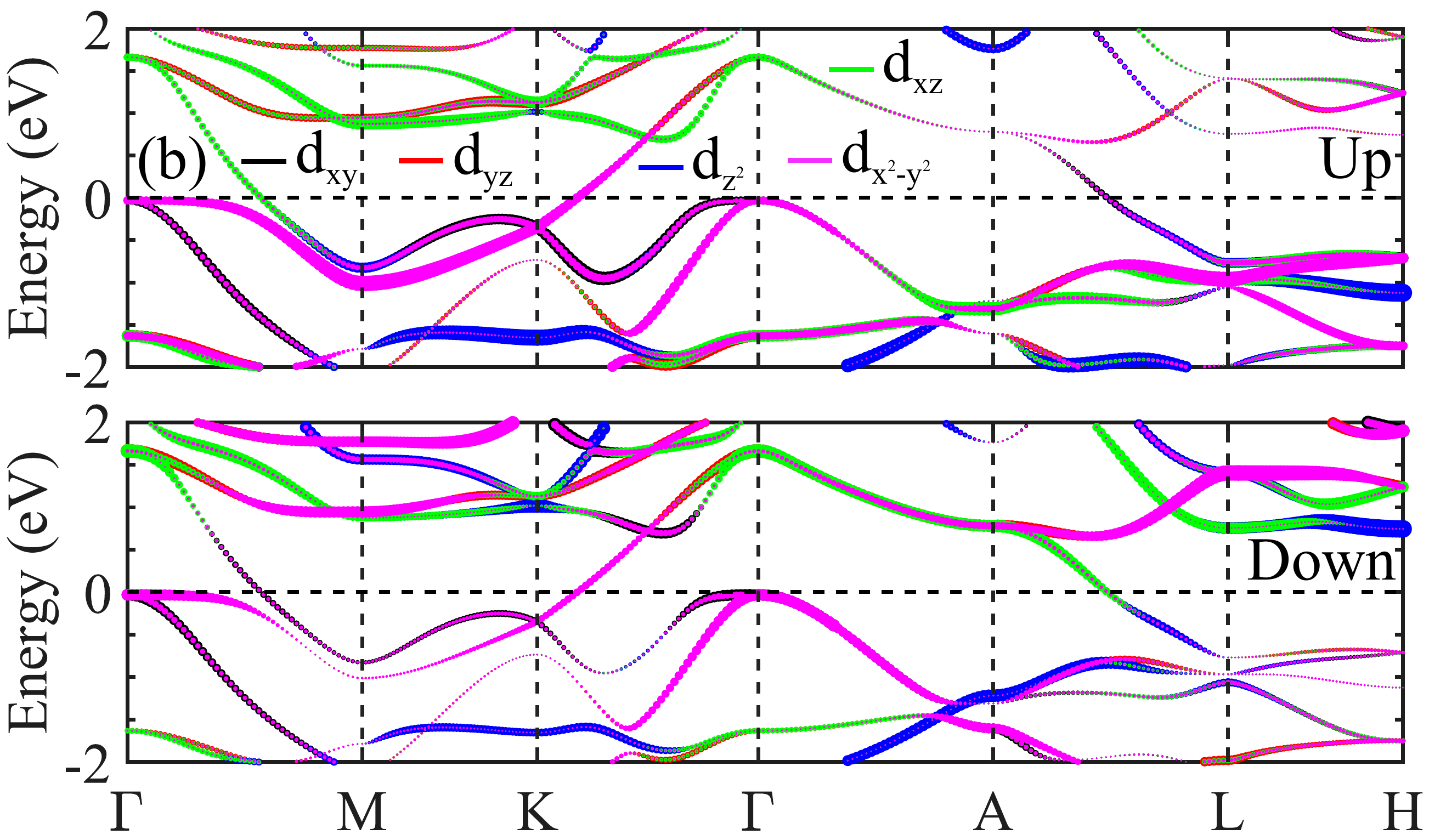}
\includegraphics[width=.95\linewidth]{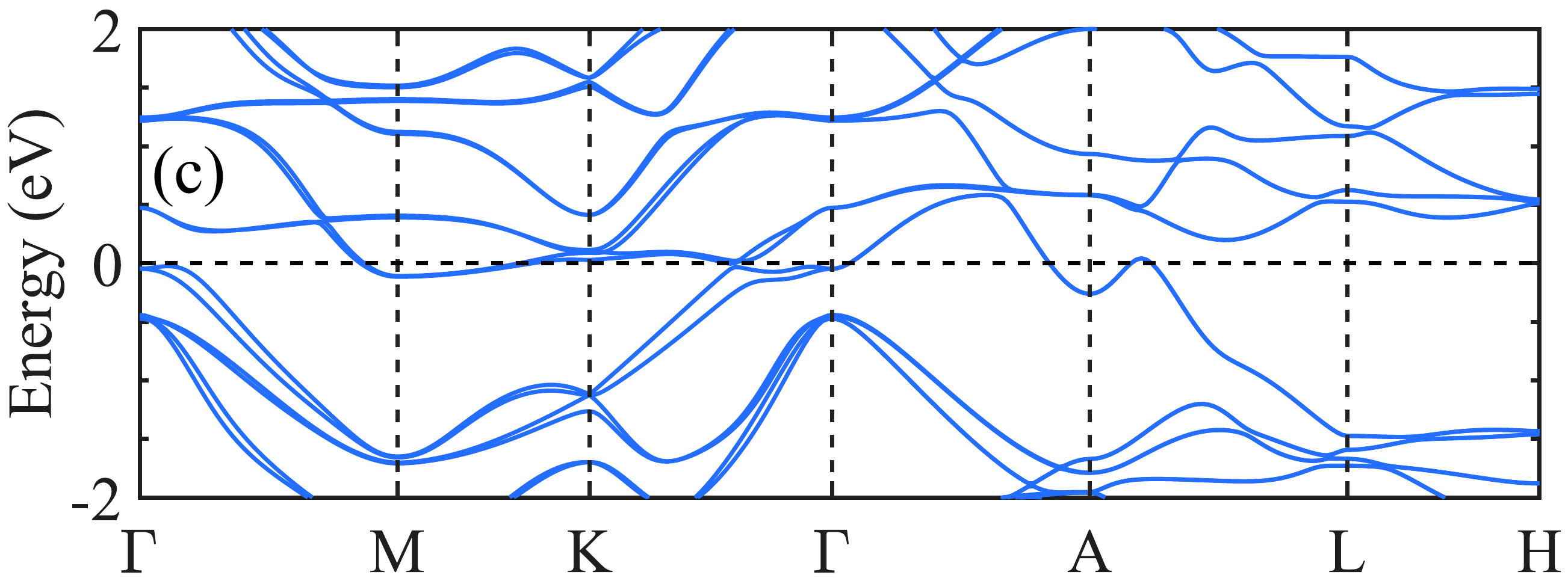}
\caption{\label{fig:band} 
The d-orbital resolved bandstructures (without SOC) of bulk CrSb when the electron number is 
(a) 22 (equilibrium) and (b) 21. 
For both (a) and (b), the top panel is for spin up, and the bottom panel is for spin down.
The colors indicate the different d-orbitals, as indicated by the legends.
The line thicknesses indicate the relative weights.
(c) The bandstructure of CrSb in equilibrium with SOC.
}
\end{figure}

\section{Results and Discussion} 
\label{sec:results}
In Fig. \ref{fig:band}, the electronic bandstructure of bulk CrSb is shown.
Fig. \ref{fig:band}(a) is the d-orbital resolved bandstructure for a Cr atom in equilibrium in the absence of SOC.
The colors denote different orbitals as indicated by the legends, 
and the line thicknesses denote the relative occupations.
The spin-up bands are shown in the top panel, and the spin-down bands are shown in the bottom panel.
Fig. \ref{fig:band}(c) shows the equilibrium bandstructure in the presence of SOC.
\begin{figure}[H]
\centering
\includegraphics[width=.8\linewidth]{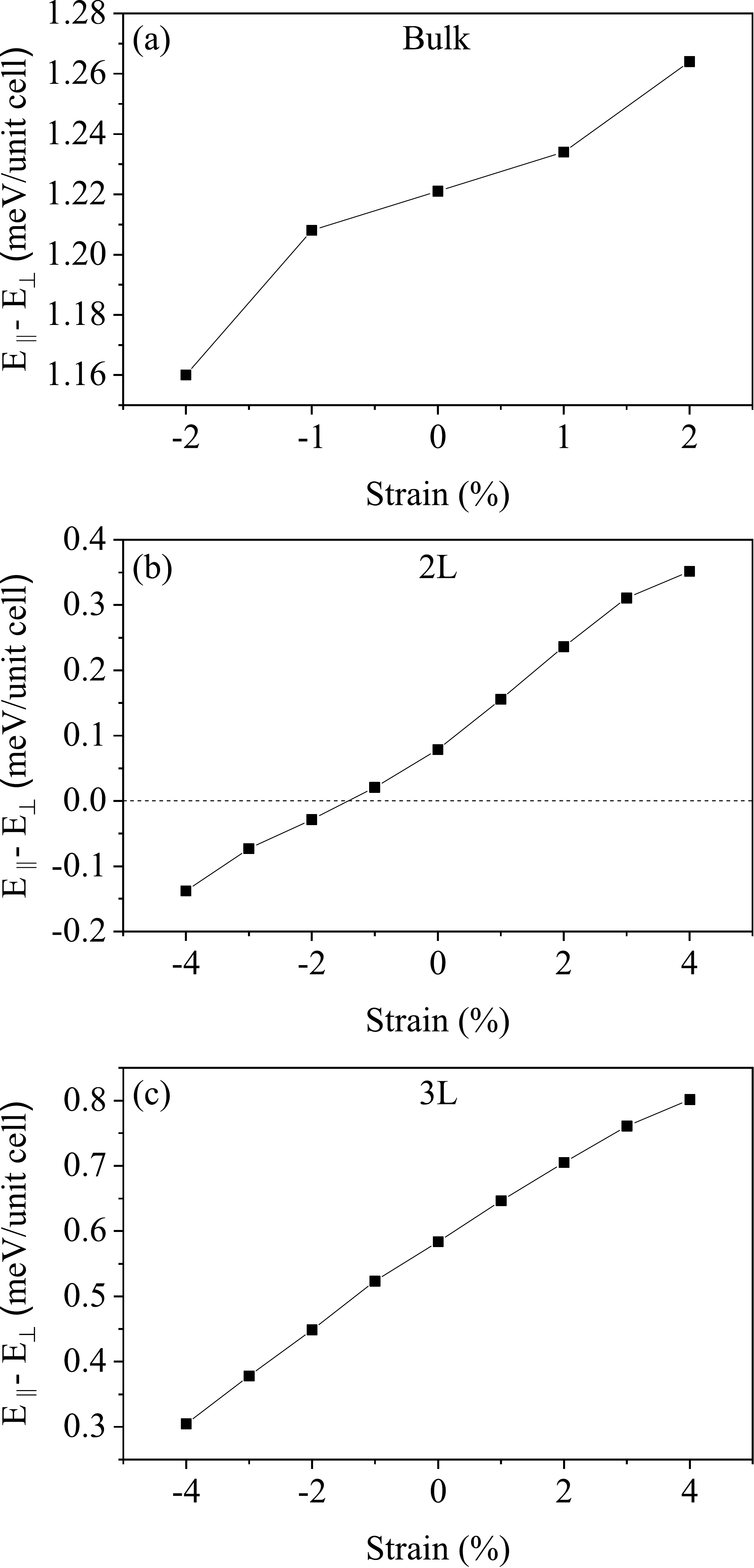}
\caption{\label{fig:strain} 
MAE as a function of applied strain for (a) the bulk crystal
and for the asymmetric (b) 1.1 nm and (c) 1.6 nm thin films.
}
\end{figure}

The effect of spin-orbit coupling (SOC) on the bandstructure is relatively large.
A comparison of Fig. \ref{fig:band}(a) to \ref{fig:band}(c)
shows that the SOC breaks the spin degeneracy throughout much of the Brillouin zone and creates
anti-crossings at a number of band-crossing points. 
The largest band splitting occurs at $\Gamma$. 
The two bands that touch at 0.2 eV in the absence of SOC 
are split by $\sim 0.5$ eV and the hole like band is pushed down below $E_F$. 
%

%%MAE versus strain - magnetostriction coefficient%%%%%%%%%%%%%
Fig. \ref{fig:strain}(a) shows $E_{\rm MAE}$ plotted as a function of applied strain for bulk CrSb.
The value at zero strain is $E_{\rm MAE} = 1.2$ meV/u.c.
The positive sign of the MAE means that the N\'eel vector aligns along the 
c-axis (out-of-plane) independent of the strain.
The monotonic increase in the MAE indicates that CrSb behaves like a magnet with a negative magnetostriction coefficient,
since the tensile strain favors out-of-plane anisotropy.
The magnetostriction coefficient ($\lambda_s$) is defined as
%% magnetostriction coefficient 
\begin{equation}
\lambda_s (ppm) = - \frac{2K_{me}(1 - v^2)}{3E \varepsilon},
\label{eq:coefficient}
\end{equation}
where $v$, $E$, and $\varepsilon$ are the Possion's ratio (0.288), Young's modulus (78.3 GPa), and strain, 
respectively \cite{lamda}.
The magnetoelastic anisotropy constant, 
$K_{me}$, is calculated from the difference between two MAEs with and without strain 
(i.e., $E_{\rm MAE}(\varepsilon) - E_{\rm MAE}(0)$).
The parameters $v$ and $E$ are taken from a previous study \cite{CrSb_mech}.
The calculated $\lambda_s$ for small strain (between $-1$ $\%$ to $1$ $\%$) is $-$19.8 ppm.
CrSb has a negative value of $\lambda_s$, and the magnitude of $\lambda_s$ 
is similar to that of MnNi and MnPd \cite{Mn_lambda}.
We also define a strain coefficient as $\alpha_\vep = dE_{\rm MAE}/d\vep$ evaluated at $\vep = 0$.
The value for bulk CrSb is $\alpha_\vep = 0.013$ meV/\%strain.

\begin{figure}[H]
\centering
\includegraphics[width=.8\linewidth]{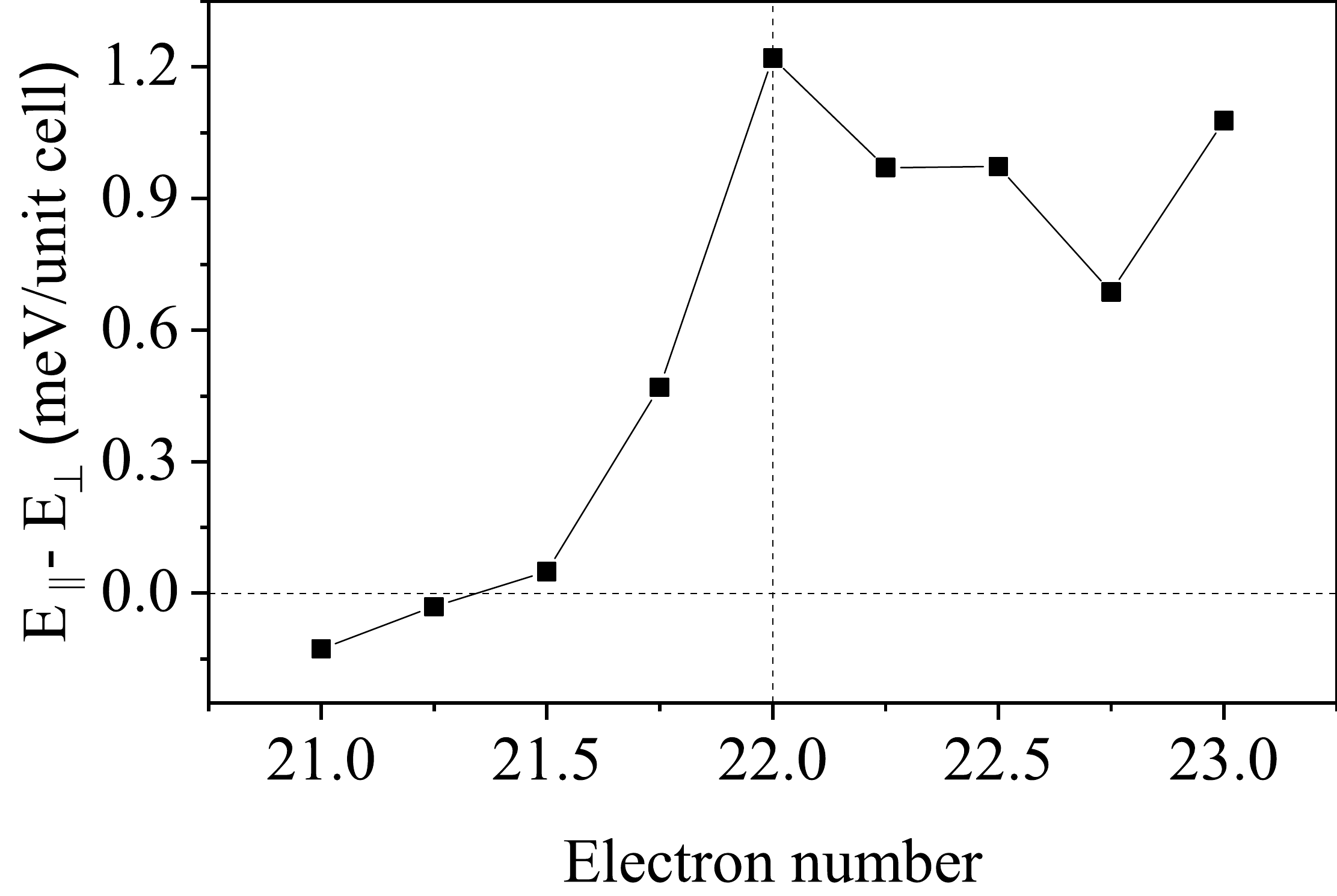}
\caption{\label{fig:filling} 
MAE of bulk crystal versus the number of electrons in the unit cell.
}
\end{figure}

%%MAE versus electron filling - explaining the sign of MAE using the second order perturbation theory
The response of the bulk MAE as a function of the electron number in the unit cell is shown in
Fig. \ref{fig:filling}.
In equilibrium, the unit cell has 22 valence electrons, which is denoted by the vertical line in the figure.
The MAE decreases most rapidly when the CrSb is depleted, and
it changes sign when the hole doping reaches 0.75/u.c.
For electron depletion, the filling coefficient, defined as $\alpha_n = dE_{\rm MAE}/dn$,
is 2.92 meV, where $n$ is the electron number per unit cell.

To understand the physical origin of the transition due to charge depletion,
we consider
the d-orbital resolved band structures for a Cr atom 
plotted for 2 different electron numbers in the unit cell,
22 (equilibrium) in Fig. \ref{fig:band}(a) and 21 in Fig. \ref{fig:band}(b).
Within second-order perturbation theory, the MAE is approximately expressed as\cite{perturb_MAE}
%% second-order perturbation equation
\begin{equation}
\begin{aligned}
MAE &\propto \xi^2 \sum_{o,u}\tfrac{\abs{\mel{\Psi^{\uparrow}_o}{\hat{L}_z}{\Psi^{\uparrow}_u}}^2 - \abs{\mel{\Psi^{\uparrow}_o}{\hat{L}_{x(y)}}{\Psi^{\uparrow}_u}}^2}{E^{\uparrow}_u - E^{\uparrow}_o} \\
&+\xi^2 \sum_{o,u}\tfrac{\abs{\mel{\Psi^{\downarrow}_o}{\hat{L}_z}{\Psi^{\downarrow}_u}}^2 - \abs{\mel{\Psi^{\downarrow}_o}{\hat{L}_{x(y)}}{\Psi^{\downarrow}_u}}^2}{E^{\downarrow}_u - E^{\downarrow}_o} \\ 
&+ \xi^2 \sum_{o,u}\tfrac{\abs{\mel{\Psi^{\uparrow}_o}{\hat{L}_{x(y)}}{\Psi^{\downarrow}_u}}^2 - \abs{\mel{\Psi^{\uparrow}_o}{\hat{L}_z}{\Psi^{\downarrow}_u}}^2}{E^{\downarrow}_u - E^{\uparrow}_o},
\label{eq:perturb}
\end{aligned}
\end{equation}
where ($\Psi_u$)$\Psi_o$, ($E_u$)$E_o$, and $\xi$ are the (un)occupied states, 
(un)occupied eigenvalues, and the spin-orbit coupling constant, respectively.
$\hat{L}_z$ and $\hat{L}_{x(y)}$ are the out-of plane and in-plane components 
of the orbital angular momentum operator, 
and $\uparrow$ and $\downarrow$ denote spin-up and spin-down.
The non-zero matrix elements in the Eq. (\ref{eq:perturb}) are 
$\mel{d_{xz}}{\hat{L}_z}{d_{yz}}$, 
$\mel{d_{x^2-y^2}}{\hat{L}_z}{d_{xy}}$, 
$\mel{d_{z^2}}{\hat{L}_x}{d_{yz}}$, 
$\mel{d_{xy}}{\hat{L}_x}{d_{xz}}$, 
$\mel{d_{x^2-y^2}}{\hat{L}_x}{d_{yz}}$, 
$\mel{d_{z^2}}{\hat{L}_y}{d_{xz}}$, 
$\mel{d_{xy}}{\hat{L}_y}{d_{yz}}$, and 
$\mel{d_{x^2-y^2}}{\hat{L}_y}{d_{xz}}$.
The largest contributions to Eq. (\ref{eq:perturb}) come from pairs of nearly degenerate occupied and unoccupied states 
near the Fermi level.
In equilibrium (Fig. \ref{fig:band}(a)), the main contributions to the MAE come from the spin-orbit coupling between 
occupied $d_{xz}^\uparrow$ and unoccupied $d_{yz}^\uparrow$ 
states at the $K$ point and near 
the $\Gamma$ point, 
and between occupied $d_{xy}^\downarrow$ and unoccupied $d_{x^2-y^2}^\downarrow$ states near the $\Gamma$ point.
All of these states couple through the $\hat{L}_z$ operator, which results in the positive MAE value (out-of-plane anisotropy).
When CrSb is depleted (see Fig. \ref{fig:band}(b)),
the entire band structure moves upward so that the main contributor states 
of the perpendicular anisotropy become unoccupied. 
This reduces the value of the MAE, and eventually reverses the sign for $n=21.25$.
%

%In FMs, the electron number can be altered by chemical doping, and a theoretical
%analysis of Fe alloyed with Co or Cr was recently described in Ref. [\onlinecite{Tsymbal_VCMA_PRB17}].
%
%However, the chemistry of substitutional hole doping of CrSb with either V or Ti would appear 
%to be more complex then that of FM alloying.
%
%While VSb exists in the same hexagonal crystalline structure as CrSb, that structure is
%higly unstable with an energy above the hull of 0.232 eV \cite{MPDatabase,MPD_Jain2013}. 
%
%All crystalline forms of VSb are non-magnetic except for V$_3$Sb$_2$ which has a 
%relatively small magnetic moment.
%
%TiSb exists in two hexagonal crystal structures. 
%
%The ground state structure is non-magnetic,
%and the unstable structure is FM with a small magnetic moment. 
%
%All other crystalline forms of VSb and TiSb are non-magnetic.
%
%Thus, substitutional doping with V or Ti may provide holes, 
%but it will also likely dilute the magnetic moment of the Cr lattice.
%

\begin{figure}[H]
\centering
\includegraphics[width=1\linewidth]{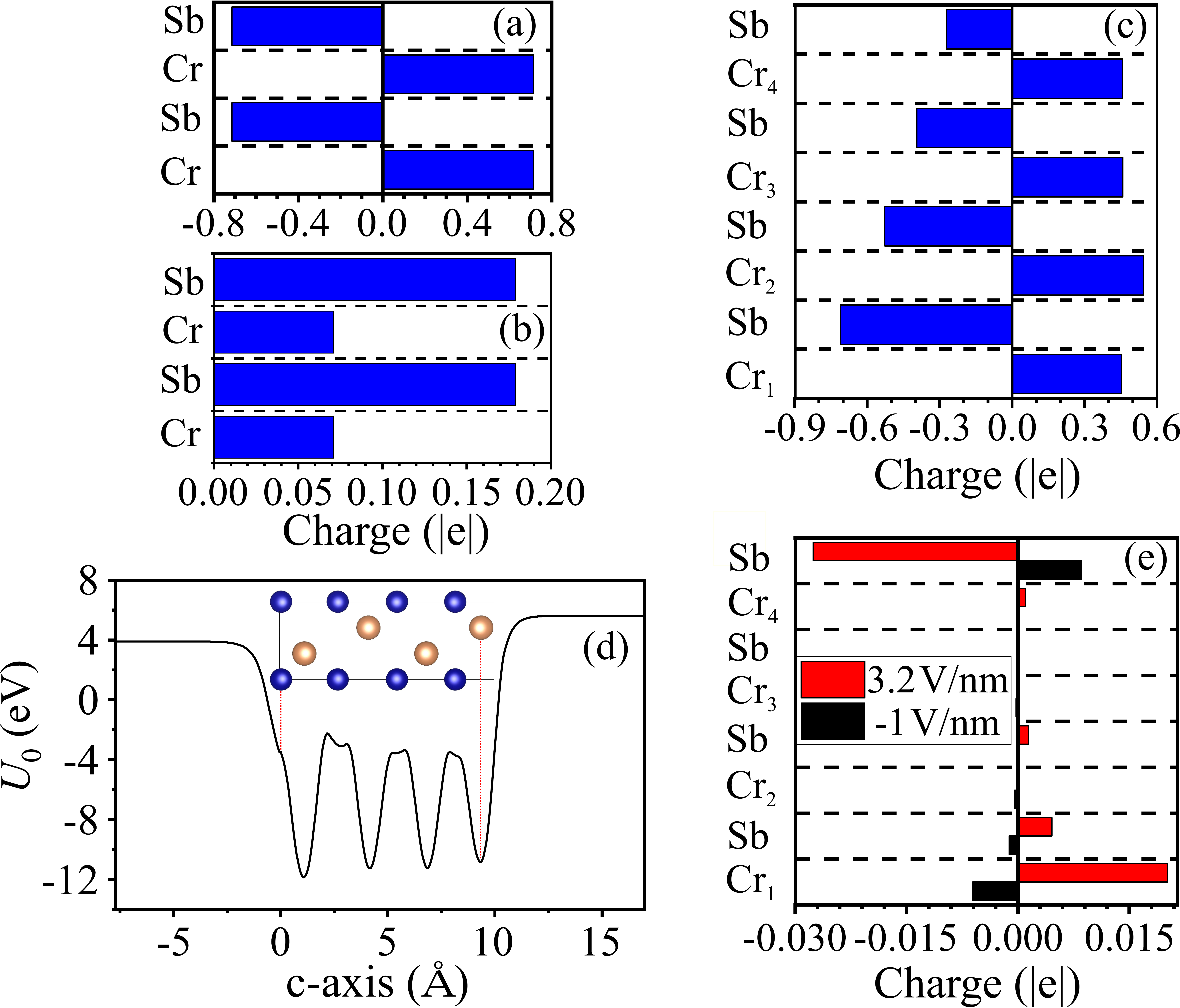}
\caption{\label{fig:charge} 
The net electronic charges on the Cr and Sb atoms, calculated from the Bader charges, 
in units of $|e|$, of (a) bulk and (c) 1.1 nm thin film CrSb in equilibrium. 
(b) Change in net electronic bulk charge due to hole doping of 0.5 holes / unit cell.
(d) The planar-averaged Hartree potential of the 1.1 nm thin film CrSb in equilibrium. 
(e) Change in the net electronic charges induced by the electric field (indicated in the legend) 
with the reference charge taken from equilibrium charges shown in (c).
Note that a net positive electronic charge corresponds to a depletion of the electron density.
}
\end{figure}

Below, we will
compare the sensitivity of the bulk crystal MAE to electron filling 
with the sensitivity of the thin-film MAE to applied electric field. 
Such a comparison requires a common metric based on a common physical quantity that governs the MAE.
Assuming that the common driving mechanism is the population change of the magnetic Cr atoms \cite{Tsymbal_VCMA_PRB17},
we determine a slightly different parameter,
\begin{equation}
\alpha_{\bar{n}_{\rm Cr}} = dE_{\rm MAE}/d\nCr ,
\label{eq:alpha_n_def}
\end{equation}
where $\nCr$ is the average electron number on the Cr atoms as determined from the Bader charges.
Due to the strongly ionic nature of the Cr-Sb bond, only $\sim 1/3$ of the hole doping goes
to the Cr sublattice. 
The equilibrium charge resulting from a transfer of $\sim 0.7$ electrons from the Cr atom to the Sb atom
is shown in Fig. \ref{fig:charge}(a), and the change in charge at a filling of -0.5 electrons / u.c.
is shown in Fig. \ref{fig:charge}(b).
The filling of -0.5 electrons / u.c. corresponds to -0.07 electrons / Cr atom.
Using the values from Fig. \ref{fig:charge}(b), for hole doping, 
$\anCr = 16.2$ meV.

%%MAE versus electric field - VCMA coefficient%%%%%%%%%%%%%%%%%
We now consider thin-film slabs with
thicknesses of 2 and 3 unit cells corresponding to 1.1 nm and 1.6 nm, respectively.
%%% Need to ask him%%%%%%%%%%
%
For the thinner slab, the cohesive energies are calculated for a 7 atomic layer structure
of alternating Cr and Sb layers terminated on both ends with a Cr layer and for
a 8 atomic layer structure (2 unit cells) terminated on one end with Cr and on the other with Sb.
The cohesive energies of the 7 and 8 atomic layer structures are -3.328 eV and -3.517 eV, respectively.
For the thicker slab, cohesive energies are calculated for an 11 atomic layer structure and a 12 atomic layer 
structure, and the cohesive energies are -3.522 eV and -3.636 eV, respectively.
Thus, the slab strucures with an integer number of unit cells such that one face is a Cr layer
and the opposing face is an Sb layer are the most stable, 
and they are the ones that we will consider first.
We will refer to these structures as asymmetric slabs.

In these asymmetric slabs,
inversion symmetry is broken, since one end is terminated with a Cr layer and the other end is terminated
with a Sb layer. 
Thus, these thin films are also Janus structures.
The two unit cell asymmetric thin-film is shown in Fig. \ref{fig:structure}(b). 
Below, results for strain and VCMA coefficients are presented for both the 2 and 3 unit cell asymmetric thin-films,
and the in-depth microscopic analysis of the charge, magnetic moments, and electronic structure 
%as shown in Figs. \ref{fig:structure} and \ref{fig:slabband} 
focuses on the 2 unit cell asymmetric thin-film.

In the asymmetric thin films,
there is a net polarization of electron charge between the 
positively charged Cr layer on the bottom and the negatively charged Sb layer on the top. 
The excess charges on each atom, as determined from the Bader charges, 
in the bulk and in the 2 unit-cell slab are shown in Fig. \ref{fig:charge}(a,c), 
and it is clear that, in the slab, the charge transfer is no longer balanced layer-by-layer.
The net charge polarization gives rise to a built-in electric field that 
alternates positively and negatively within the slab, but, its average value
points from the positive Cr layer on the bottom to the negative Sb layer on the top.
This built-in electric field results in a built-in potential across the slab of 1.7 eV 
as shown by the plot of the equilibrium planar averaged Hartree potential in
Fig. \ref{fig:charge}(d).

\begin{figure}[H]
\centering
\includegraphics[width=.95\linewidth]{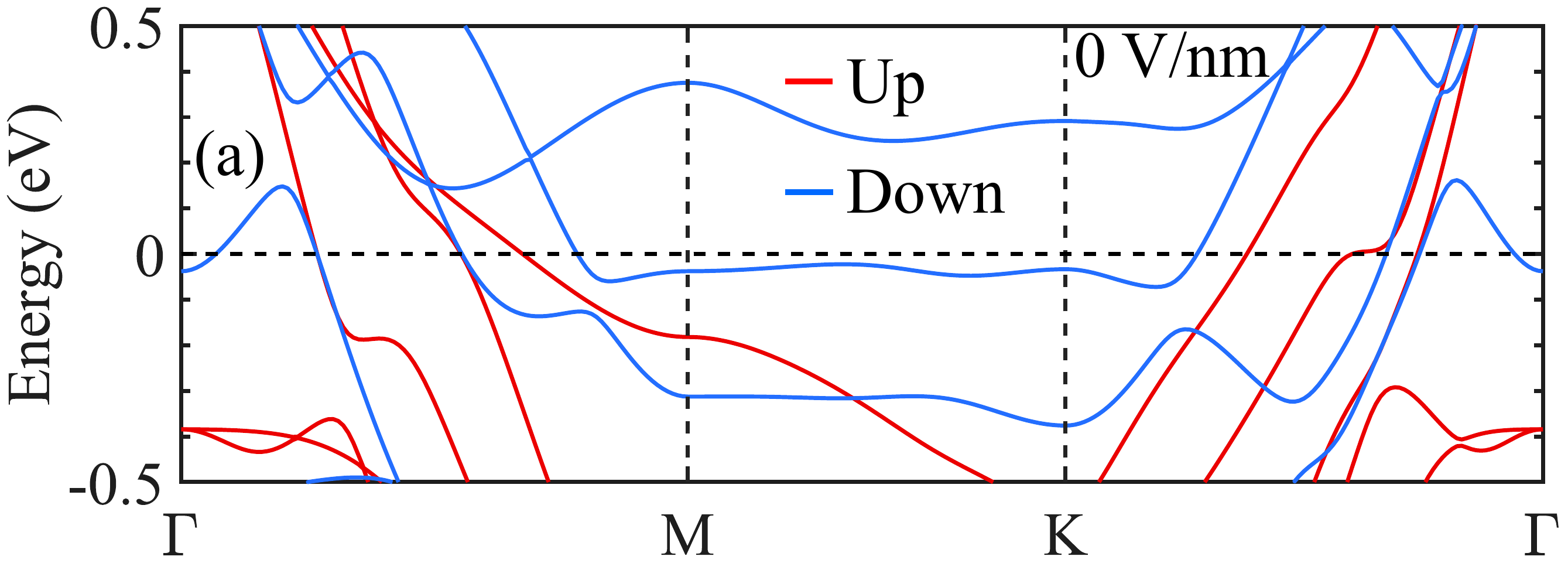}
\includegraphics[width=.95\linewidth]{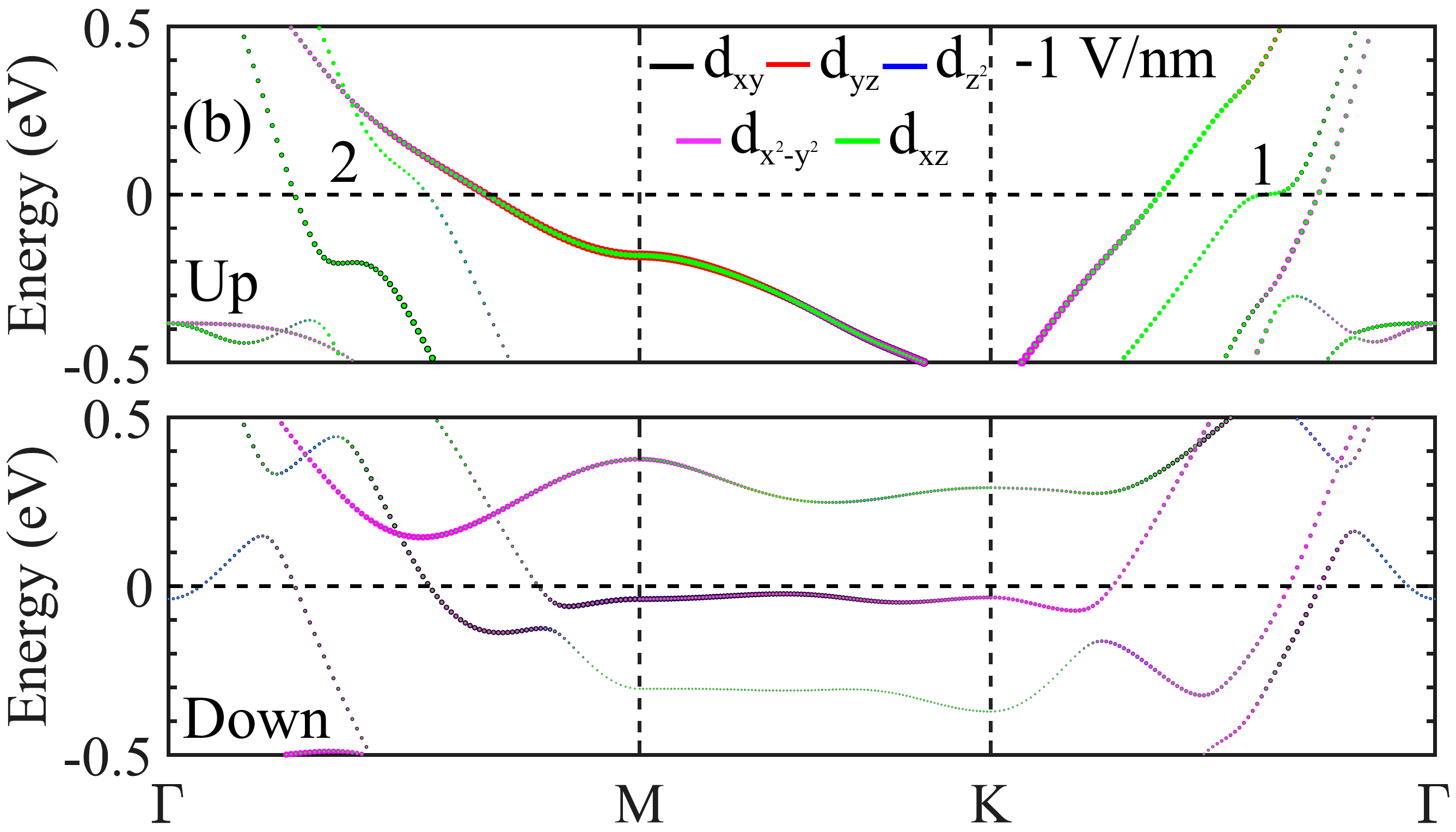}
\includegraphics[width=.95\linewidth]{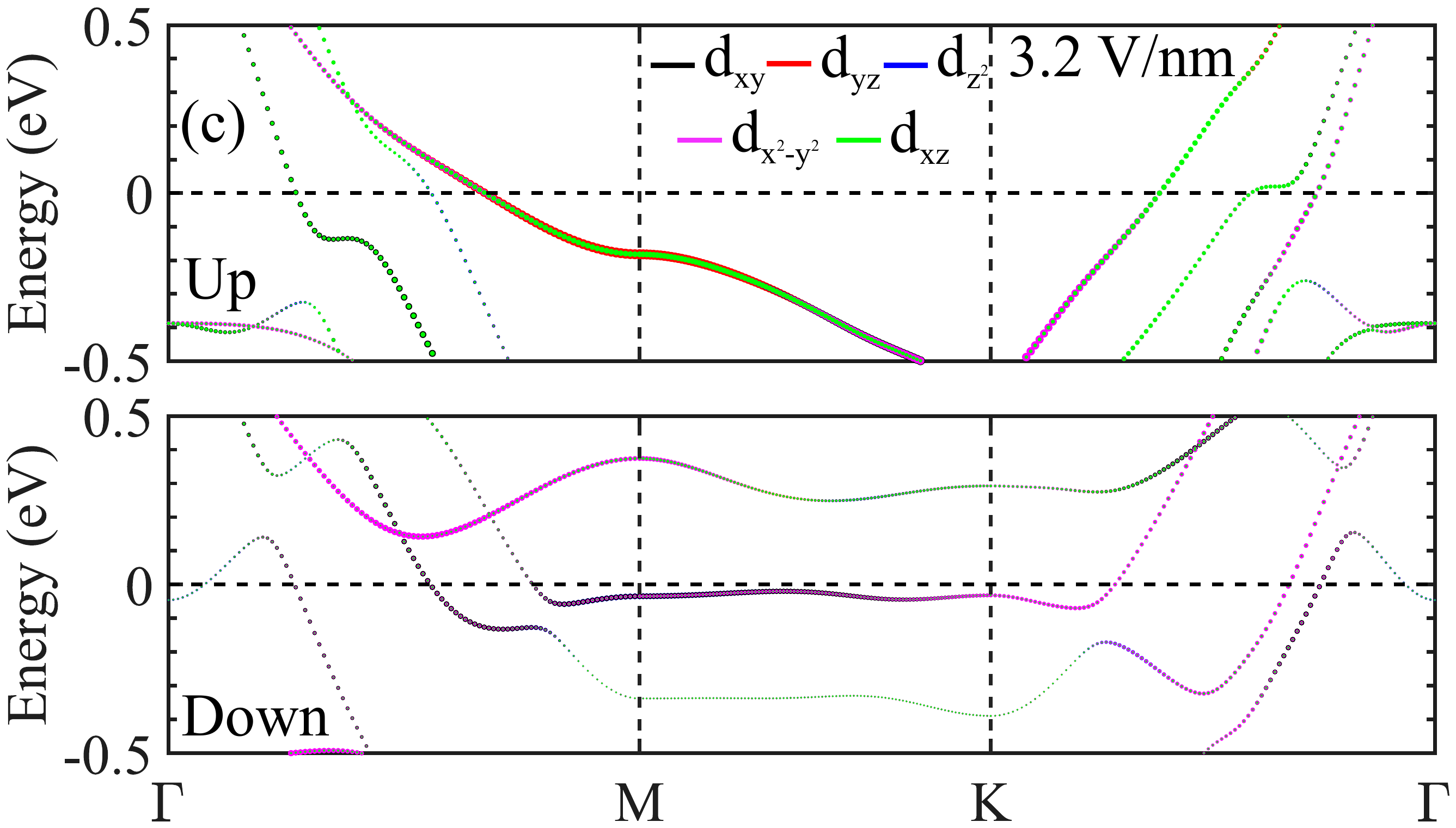}
\caption{\label{fig:slabband} 
(a) Spin resolved bandstrucure of the 1.1 nm CrSb thin-film in equilibrium.
(b,c) The d-orbital resolved bandstructures of the 1.1 nm CrSb thin-film under electric fields of 
(b) $-$1 V/nm and (c) 3.2 V/nm. 
The line colors indicate the d-orbital composition as given by the legends, 
and the line thicknesses indicate the relative weights. 
}
\end{figure}

The truncation of the bulk to a finite slab results not only in a
loss of {\em local} balance between the positive and negative charges, 
but also in a {\em global} imbalance of the magnetic moments of the Cr ions.
In other words, the cancellation of magnetic moments between alternating layers of Cr is no longer exact,
and a small net magnetic moment exists in the slab.
The magnetic moments on each Cr atom are listed in the `$\mu_{\rm Cr,Sb}$' column of 
Table \ref{table:SOC_magmom} with the numbering 
of the Cr atoms corresponding to that shown in Fig. \ref{fig:structure}(b).
The magnitudes of the magnetic moments in the Cr layers monotonically decrease from bottom to top
as the Cr atoms approach the Sb terminated end of the slab.
The breaking of the spin degeneracy is readily apparent in the bandstructure of the slab shown in 
Fig. \ref{fig:slabband}(a). 
The degeneracy between the up-spin and down-spin bands is broken, and
the CrSb slab has become a ferrimagnet (FiM).
The breaking of the spin degeneracy of the AFM states in a finite slab of a layered AFM 
is explained by a simple chain model with different end terminations
described in [\onlinecite{NDjavid_PRB20_note}].
The wavefunctions of a pair of degenerate AFM states are weighted differently
on alternate atoms of the magnetic lattice. 
Thus, the coupling of the two states to an end atom is different, and this different
coupling breaks the degeneracy of the two states.
%
%Since any real slab will rest on a substrate and also likely be capped with a different material, 
%it is expected that all real thin slabs of bulk layered AFMs will be FiMs rather than AFMs.

%
\begin{table}[H]
\centering
\caption{Magnetic moment, in units of $\mu_B$, 
for each Cr atom in the bulk and thin films of Fig. \ref{fig:structure}
with spin-orbit coupling.
The indices on the Cr atoms correspond to those in Fig. \ref{fig:structure}.
$\mu_{\rm Cr,Sb}$ corresponds to the thin film of Fig. \ref{fig:structure}(b)
with an integer number of unit cells,
and $\mu_{\rm Cr,Cr}$ corresponds to the thin film of Fig. \ref{fig:structure}(c)
in which the top Sb layer is removed.
}
\label{table:SOC_magmom}
{\setlength{\tabcolsep}{12pt}
\begin{tabular}{c c c c}
\hline\hline
Atom   & $\mu_{\rm bulk}$ & $\mu_{\rm Cr,Sb}$ & $\mu_{Cr,Cr}$ \\ \hline
Cr$_1$ & 3.035 & 3.763 & 3.905 \\
Cr$_2$ & $-$3.035 & $-$3.190 & $-$2.994 \\
Cr$_3$ & & 3.116.& 2.994 \\
Cr$_4$ & & $-$2.889.& $-$3.904 \\
\hline\hline
\end{tabular}
}
\end{table}

A third result of truncating the bulk to a slab is that the MAE decreases.
For the 1.1 nm slab, the MAE is reduced by a factor of 15 from
1.2 meV/u.c. to 0.079 meV/u.c. (0.17 erg/cm$^2$), 
and for the 1.6 nm slab, the MAE is reduced by a factor of 2.1 to 0.58 meV/u.c. (1.85 erg/cm$^2$).
Also, the sensitivity of the MAE to strain increases.
The strain coefficients of the 1.1 and 1.6 nm slabs ($\alpha_\vep$) increase 
from 0.013 meV/\%strain in the bulk 
to 0.068 meV/\%strain and 0.062 meV/\%strain, respectively, 
where the energies are per bulk unit cell (2 Cr atoms).
The combined result of the reduced MAE and increased strain coefficient
is that a 1.5\% uniaxial compressive strain along [1000] direction 
in the 1.1 nm thin film causes a $90^\circ$ rotation of the 
N\'{e}el vector from out-of-plane to in-plane as shown in Fig. \ref{fig:strain}(b).

\begin{figure}[H]
\centering
\includegraphics[width=.85\linewidth]{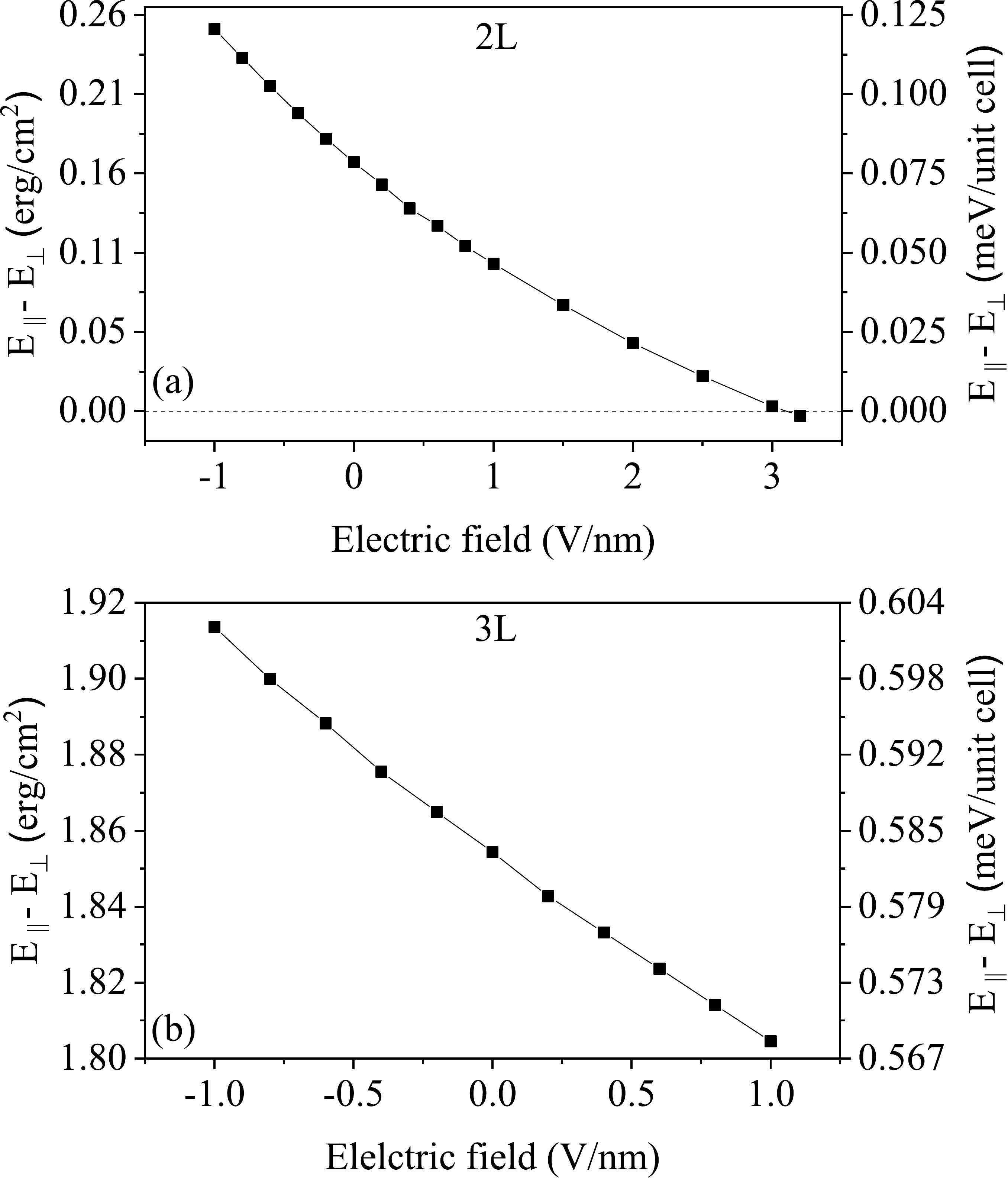}
\caption{\label{fig:field} 
MAEs of (a) 1.1 nm and (b) 1.6 nm films as a function of applied electric field.
}
\end{figure}

A fourth result is that the MAE also becomes sensitive to an external electric field 
as shown in Fig. \ref{fig:field}.
In other words, the thin slab exhibits intrinsic VCMA.
Typically VCMA is found when a magnetic layer is placed in contact with a heavy-metal layer that provides
SOC.
However, the Sb layers provide large SOC, and the terminating Sb layer serves as the HM layer, such that the
CrSb slab has intrinsic VCMA.
The MAE decreases linearly as the electric field is increased, and,
for the 1.1 nm slab, it changes sign at 3.2 V/nm,
which indicates that the N\'eel vector rotates 90$^\circ$ 
from out-of-plane to in-plane.

\begin{figure}[H]
\centering
\includegraphics[width=1\linewidth]{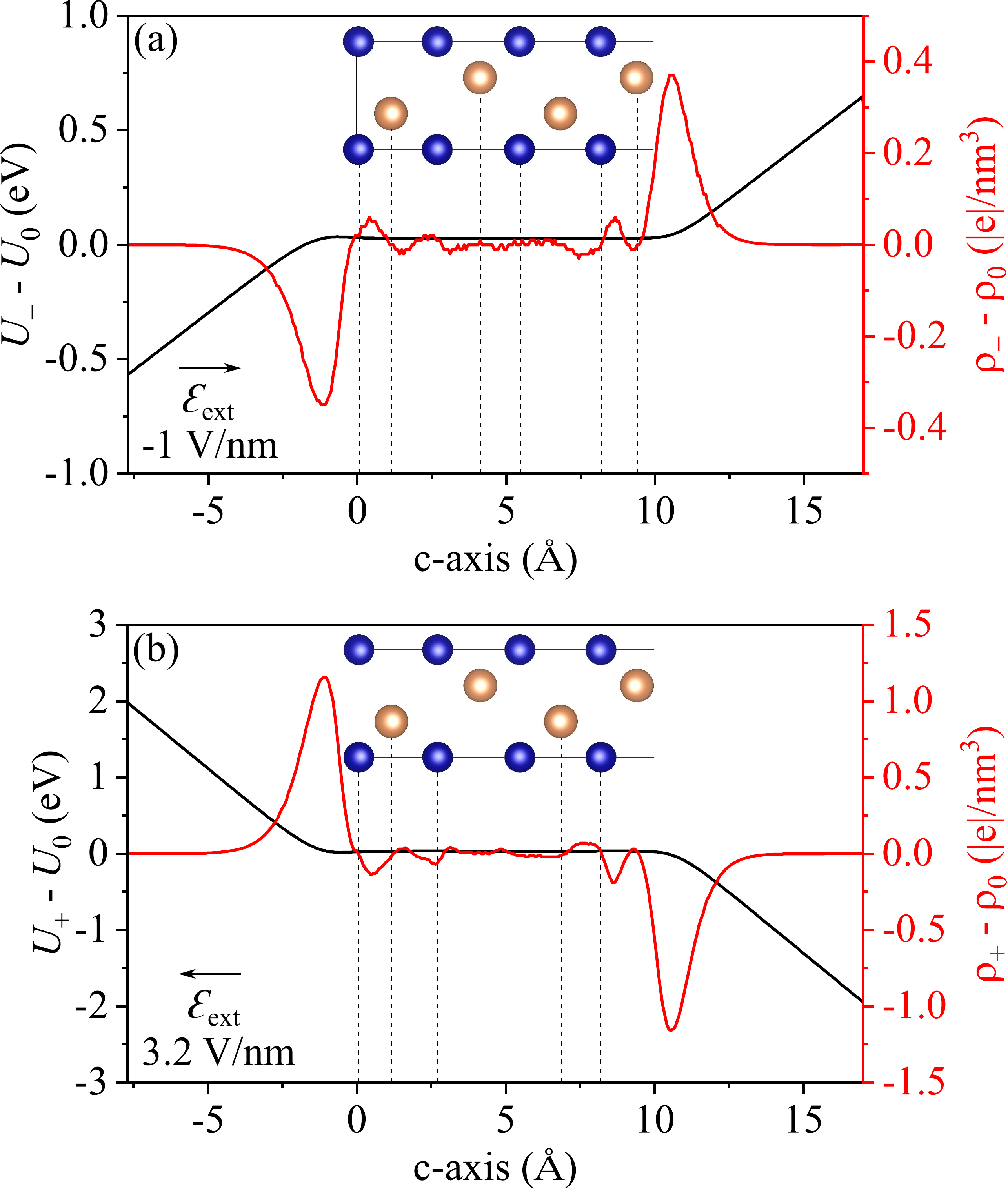}
\caption{\label{fig:pot_and_cha} 
The change in the planar-averaged electronic charge (red) and planar-averaged Hartree energy (black) 
of 1.1 nm thin film CrSb with an electric field of (a) -1 V/nm and (b) 3.2 V/nm. 
The arrow indicates the direction of applied electric field.
Note that a positive electronic charge corresponds to a depletion of the electron density.
}
\end{figure}

The standard metric describing the sensitivity of the MAE to the applied electric field is given by
the VCMA coefficient defined as 
\begin{equation}
\beta = \frac{d E^s_{\rm MAE}}{d\sE_I} = \frac{dE^s_{\rm MAE}}{d\sE_{\rm ext}/\ep_I} ,
\label{eq:VCMA_coeff}
\end{equation}
where $E^s_{\rm MAE}$ is the MAE per unit area of the slab, 
$\sE_{\rm ext}$ is the external electric field and
$\sE_{I}$ and $\ep_{I}$ are the electric field and the relative dielectric constant in the insulator, repectively.
In a typical experiment, the insulator would be an oxide layer with $\ep_I > 1$.
In our case, the insulator is the vacuum, so that $\ep_I = 1$ and $\sE_I = \sE_{\rm ext}$. 
The values of $\beta$, evaluated at $\sE_{\rm ext}=0$ V/nm, 
are $-$76.4 fJV$^{-1}$m$^{-1}$ for the 1.1 nm film and $-$55.3 fJV$^{-1}$m$^{-1}$ for the 1.6 nm slab.
%

%\textcolor{blue}{
The calculated magnitudes of $\beta$ are comparable with many of the experimentally measured values of $\beta$ 
from heavy metal/FM/MgO heterostructures,
\cite{shiota2017reduction,kawabe2017electric,skowronski2015underlayer,
skowronski2015perpendicular,li2017enhancement,
nozaki2014magnetization,hibino2015electric,VCMA_Exp_SciRep17}
with magnitudes ranging from 65 {\fJVm} in a Ta/Co$_{0.25}$Fe$_{0.55}$B$_{0.20}$/MgO structure 
\cite{shiota2017reduction}
to 139 {\fJVm} in a Ru/Co$_2$FeAl/MgO structure. \cite{VCMA_Exp_SciRep17}
However, they are considerably smaller than the values of
370 {\fJVm} and 1043 fJV$^{-1}$m$^{-1}$ 
measured in Cr/Fe/MgO and Ir/CoFeB/MgO structures, 
respectively. \cite{2017_VCMA_Cr_Fe_MgO_ScRep,kato2018giant}
A lengthy table of experimental values compiled from the literature is provided 
in a recent review \cite{VCMA_review_Miwa_JPD18}.
Theoretical studies of AFM materials have calculated magnitudes of 
$\beta$ both larger and smaller than the values for CrSb. 
For a G-AFM FeRh/MgO bilayer, the magnitude of $\beta$ was in the range
of 130 to 360 {\fJVm} depending on the sign of the electric field, the strain,
and the termination layer of either Fe or Rh \cite{zheng2017electric}.
A value of 22.6 {\fJVm} was calculated for Pt/MnPd with vacuum as the dielectric \cite{su2020voltage}.
VCMA calculations of a MgO/MnPt/MgO slab found magnitudes of 170 {\fJVm} and 70 {\fJVm} 
depending on whether the MnPt
layer was terminated on both ends with either Mn or Pt, respectively \cite{chang2020voltage}.
%}

To understand which Cr atoms contribute to the VCMA effect, 
we inspect the change in charge of the Cr atoms in response to the applied field.
The change in charge with applied electric field, plotted in Figs. \ref{fig:charge}(e)
and \ref{fig:pot_and_cha}, shows that,
among the Cr atoms, the only significant change in charge occurs on Cr$_1$.
This is to be expected, since the electric field is screened by the first and 
last atomic layers from the interior of the CrSb metal.
The first atomic layer is Cr and the last is Sb, so that an applied field
creates a net change in charge on the Cr sublattice with the charge being transferred to the
Sb sublattice.

The sensitivities of the MAE of the thin film slabs to 
the average charging of the Cr atoms due to an applied electric field
is comparable to
the sensitivity of the bulk MAE to the filling of the Cr atoms.
% 
%Define $\bar{n}_{\rm Cr}$ as the average number of electrons on the Cr atoms,
%as determined from the Bader charges, and
%$E_{\rm MAE}$ as the MAE per unit cell or per two Cr atoms.
%
The values of
$\anCr$
evaluated at $\sE_{\rm ext}=0$ are 19.7 meV and 13.9 meV 
for the 2 layer and 3 layer slabs, respectively.  
For comparison, the bulk value for hole doping is $\anCr = 16.2$ meV.
Thus, the sensitivity of the MAE per unit cell to the change in the average charge on the Cr atoms
lies in the range of 13 - 20 meV for both of the slabs and the bulk.
To elucidate the physical origin of switching mechanism in the two layer slab, 
the d-orbital resolved band structures for the Cr${_1}$ atom under different electric fields 
are plotted in Fig. \ref{fig:slabband}(b,c).
At the electric field of $-$1 V/nm, the major contribution of the perpendicular anisotropy 
comes from the spin-orbit coupling between the unoccupied $d_{xz}^\uparrow$ 
and occupied $d_{x^2 - y^2}^\downarrow$ states through $\hat{L}_y$ operator in the region 1.
Although the states in the region 1 are coupled through the in-plane angular momentum operator, 
the contribution of the MAE is positive since they are from different spin channels.
In the region 2, the occupied $d_{xy}^\uparrow$ states are coupled with unoccupied $d_{xz}^\uparrow$ states through $\hat{L}_x$, 
which contributes the in-plane anisotropy.
As the external field increases (see Fig. \ref{fig:slabband} (c)), 
the unoccupied $d_{xz}^\uparrow$ states in the region 1 move away from the Fermi level by 18 meV, 
which results in the reduction of the out-of-plane anisotropy, since the denominator in the 
Eq. \ref{eq:perturb} increases.
In addition, in the region 2, the occupied $d_{xy}^\uparrow$ states move closer to the Fermi energy by 67 meV 
as the electric field increases so that the in-plane anisotropy increases.
%
%

%\textcolor{blue}{
Previous studies have shown that 
the choice of surface termination of a FM layer can alter, or even change the sign
of the 
MAE \cite{2011_MAE_Fe-MgO_Co-MgO_Shin_PRB,2014_iPMA_Fe-MgO_Yokoyama_APL,zheng2017electric,chang2020voltage}. 
Thin film CrSb is no exception.
%
%Furthermore, the effect of surface termination in CrSb is even more pronounced than those that
%have been observed in other materials such as FeRh and MnPt.
%
%We attribute this extreme sensitivity to the large ionic nature of the bonding in 
%CrSb that results in large charge transfer between the Cr and Sb layers.
%
%The Bader charge analysis for the bulk crystal gives a charge transfer of 0.7 electrons from the
%Cr atom to the Sb atom as shown in Fig. \ref{fig:charge}(a).
%%
%The nature of the bonding is significantly more ionic than any other FM or AFM material that
%has been previously investigated for VCMA.
%
%Thus, it might be expected that switching the termination of the top layer from Sb to Cr
%may have a significant effect on the VCMA, and, indeed, it does. 
%
The final structure that we consider is the slab shown in Fig. \ref{fig:structure}(c).
It is identical to the 1.1 nm thin film analyzed above, except that the top Sb layer 
is removed so that the thin film becomes mirror symmetric with respect to the $x-y$ plane.
As a result, there is no net electrical dipole moment or built-in potential across the slab. 
The net magnetic moment of the slab remains zero, as shown in the last column of Table \ref{table:SOC_magmom}.
The magnetic moment is no longer compensated locally in each unit cell; 
the compensation occurs between the mirror symmetric pairs of Cr atoms.
Furthermore, the magnitude of the MAE ($-1.07$ meV / u.c) is more than an order of magnitude larger
compared to that of the asymmetric slab ($0.079$ meV / u.c), 
and it changes sign, so that in-plane alignment of the magnetic moments is preferred.
The magnitude of the MAE is similar to that of the bulk ($1.2$ meV / u.c), but with opposite sign.
Note that the value of $-1.07$ meV / u.c. is obtained by taking the total MAE of the slab and
dividing by two, since it contains two unit cells of magnetic ions even though the last Sb layer of the
top unit cell is missing. 

%Overall, the magnetic texture of the slab retains the $\mathcal{PT}$ symmetry
%of the bulk unit cell 
%where $\mathcal{P}$ and $\mathcal{T}$ are the parity and time reversal operators, respectively.
%

Applying an electric field to this symmetric slab depletes electrons from the bottom Cr layer and 
accumulates electrons on the top Cr layer, 
so that the net change in charge on the Cr layers is zero.
This is in contrast to the effect of an electric field on the antisymmetric slab where the
applied field depletes electrons from the bottom Cr layer and accumulates electrons on the top
Sb layer, with the overall effect being a net depletion of electrons on the Cr atoms.
In the symmetric slab, with no net change in charge on the Cr atoms, the MAE remains unchanged 
to 4 significant digits over the range of $-1 \le \sE_{\rm ext} \le 3$ V/nm. 
This is consistent with results of prior simulations of symmetric MgO / FM / MgO 
and MgO / AFM / MgO
structures \cite{2008_VCMA_Tsymbal_PRL,chang2020voltage}.

To estimate the sensitivity of the MAE of the symmetric slab to 
electron filling, we alter the electron number of the slab by applying a compensating
background charge, as we did in the bulk.
Since the background charge is uniformly distributed throughout the simulation domain, which includes
the vacuum region,
the majority of the compensating charge in the CrSb slab
is located on the outer two Cr layers.
Depleting 0.5 electrons from the CrSb slab results in a total reduction of 0.32 electrons from
the Cr sublattice with an average reduction of 0.080 electrons from the Cr atoms.
The MAE changes from $-1.07$ meV to $-0.7$ meV, so that $\anCr = -4.6$ meV.
The negative sign means that as electrons are removed, the in-plane orientation of the N\'eel 
vector becomes less stable.
The magnitude of $\anCr$ is a factor of 4 less than that of the asymmetric slab,
and the sign is opposite. 
Thus, the mirror symmetric slab of just 4 Cr layers has a high in-plane MAE,
similar in magnitude to that of the bulk, and it 
is relatively insensitive to filling.
Physically realizing such a structure would be challenging.

In a typical physical structure, the CrSb slab will be sandwhiched between a MgO layer on one face
and a grounded heavy-metal (HM) layer on the opposing face.
The electric field in the dielectric MgO will terminate at the CrSb where it will
accumulate or deplete charge on the first atomic layer of the CrSb as shown
at the left of Fig. \ref{fig:pot_and_cha}. 
In the physical structure, there is no corresponding charge depletion or accumulation on
the opposing face of the CrSb slab, since the HM is grounded and supplies the charge required
to screen the electric field at the MgO/CrSb interface. 
Also, there will naturally be asymmetry and a built in potential across the CrSb slab due to the proximity of 
MgO on one face and a HM on the other.
Since breaking the mirror symmetry of the slab breaks the degeneracy of the AFM states,
it is most probable that the CrSb in a MgO/CrSb/HM structure will be in a FiM state. 
For the electric field in the MgO
to significantly alter the charge on the Cr sublattice, 
the CrSb should be terminated with a Cr layer at the MgO interface.

Finally, we note that 
the parameter $\anCr$, which gives the sensitivity of the MAE to the magnetic sublattice filling, 
is simply related to the conventional VCMA parameter $\beta$ by
\begin{equation}
\anCr \approx  -\frac{2|e|}{\ep} \beta ,
\label{eq:beta_alpha}
\end{equation}
where $\ep = \ep_0 \ep_I$, is the dielectric constant of the insulator,
and
the the negative sign is consistent with the sign of the positive electric field
and the orientation of the slab in Fig. \ref{fig:structure}. 
The sign would reverse if either the field or the slab were reversed.
Equation (\ref{eq:beta_alpha}) is derived by noting that the induced charge lies
primarily on Cr$_1$, and, therefore, it can be approximated as a sheet density
of an ideal metal given by
$n_s = n_{\rm Cr_1}/A_{u.c.} \approx \nCr N_L^{\rm Cr}/A_{u.c.} \approx -\ep \sE$ 
where $A_{u.c.}$ is the area of the unit cell in the basal plane,
and $N_L^{\rm Cr}$ is the number of Cr layers in the slab.
Also, $E^s_{\rm MAE} = E_{\rm MAE}  N_{u.c.} / A_{u.c.}$ where $N_{u.c.} =  N_L^{\rm Cr}/2$.
With these relations, we can write, 
$dE_{\rm MAE}/d\nCr \approx -\frac{2|e|}{\ep} dE^s_{\rm MAE} / d\sE $,
which is Eq. (\ref{eq:beta_alpha}).
This expression slightly underestimates the magnitude of $\anCr$, since the screening is not ideal.
For example, in the asymmetric 2-layer slab, $\beta = -76.4$ {\fJVm}.
Using this value in Eq. (\ref{eq:beta_alpha}), 
gives $\anCr = 17.3$ meV,
whereas the actual value is $\anCr = 19.7$ meV.
This relationship between  $\anCr$ and $\beta$ 
assumes that the slab terminates with a magnetic layer adjacent to the dielectric,
and the one term specific to CrSb comes from the ratio $N_l^{\rm Cr}/N_{\rm u.c.}$
that was explicitly evaluated to give the factor of $2$ in Eq. (\ref{eq:beta_alpha}).
This provides a simple relationship between the conventional metric $\beta$
and the sensitivity of the MAE to the underlying driving mechanism of sublattice filling.

\vspace*{6pt}
\section{Summary and Conclusions}
\label{sec:conclusion}
The effects of strain, band filling, and electric field on the MAE of bulk and thin-film CrSb are determined
and analysed.
A new metric that describes the sensitivity of the MAE to the filling of the magnetic sublattice
provides a means to compare the effects of electric field and band filling on the MAE.
The magnitude of the bulk magnetostriction coefficient is comparable with 
those from other antiferromagnets and ferromagnets,
however the MAE is large (1.2 meV/u.c.) and its sign cannot be changed by strain for bulk material. 
For bulk CrSb, depleting the electron density by 0.75 electrons per unit cell depletes the flat, 
nearly-degenerate
d-orbital bands near the Fermi energy and causes a $90^\circ$ rotation of the N\'{e}el vector
from out-of-plane to in-plane.
Due to the significant ionic nature of the Cr-Sb bond, finite slabs are strongly affected
by end termination. 
Truncation of the bulk crystal to a thin film 
consisting of an even number of unit cells, 
such that one face is a Cr layer and the opposing face is an Sb layer,
breaks inversion symmetry, creates a large charge dipole and potential difference across the slab, 
and breaks spin degeneracy such that the CrSb slab becomes a ferrimagnet.
For the 1.1 nm (1.6 nm) slab, the MAE is reduced from 1.2 meV/u.c.
to 0.079 meV/u.c. (0.58 meV/u.c) and the strain coefficient is increased
from 0.013 meV/\%strain to 0.068 meV/\%strain (0.062 meV/\%strain). 
As a result of the reduced MAE and increased strain coefficient, 
the sign of the MAE in the 1.1 nm slab can be switched with 
1.5\% uniaxial compressive strain.
The large SOC from the Sb combined with broken inversion symmetry of the thin film results in an intrinsic
VCMA.
The calculated VCMA coefficients for the free-standing 1.1 nm and 1.6 nm thin films with vacuum as the insulator are 
$-$76.8 fJV$^{-1}$m$^{-1}$ and $-$55.6 fJV$^{-1}$m$^{-1}$, respectively.
If the CrSb slab is terminated with Cr layers on both faces, then it remains a compensated AFM,
but with the compensation occurring nonlocally between mirror symmetric Cr pairs.
The MAE changes sign so that in-plane alignment of the moments is preferred, 
the magnitude of the MAE remains large similar to that of the bulk, and it is
relatively insensitive to filling.
Finally, in a standard experimental configuration, the CrSb slab will have different end
terminations with MgO on one face and a HM on the other, 
so that the FiM state of the asymmetric slab
will be the most probable one observed experimentally. 

\vspace*{6pt}
\noindent
{\em Acknowledgements:} This work was supported as part of Spins and Heat in Nanoscale Electronic Systems (SHINES) an 
Energy Frontier Research Center funded by the U.S. Department of Energy, Office of Science, 
Basic Energy Sciences under Award {DE-SC0012670}. 
This work used the Extreme Science and Engineering Discovery
Environment (XSEDE)\cite{towns2014xsede}, which is supported by National
Science Foundation Grant No. ACI-1548562 and allocation
ID TG-DMR130081.
%\newpage

%\bibliography{bioRev}

\end{document}